\documentstyle[preprint,aps]{revtex} 

\begin{document}

\draft

\title{Effect of the vacancy interaction on the antiphase domain growth process
in a two dimensional binary alloy} \author{Marcel Porta, Carlos Frontera, Eduard
Vives and Teresa Cast\'an}

\address{ Departament d'Estructura i Constituents de la Mat\`{e}ria, Facultat de
F\'{\i}sica, \\Universitat de Barcelona, Diagonal 647, E-08028 Barcelona,
Catalonia, Spain.}

\date{\today}

\maketitle

\begin{abstract}

We have performed a Monte Carlo simulation study of the influence of diffusing
vacancies on the antiphase domain growth process in a binary alloy after a
quench through an order-disorder transition.  The problem has been modeled by
means of a Blume-Emery-Griffiths hamiltonian which biquadratic coupling
parameter $K$, controls the microscopic interactions between vacancies.  The
asymmetric term $L$ has been taken $L=0$ and the ordering dynamics has been
studied at very low temperature as function of $K$ inside the range $-0.5 \leq
K/J \leq 1.40$ (with $J>0$ being the ordering energy).  The system evolves
according to the Kawasaki dynamics which conserves the alloy concentration while
the order parameter does not.  The simulations have been performed on a
two-dimensional square lattice and the concentration has been taken so that the
system corresponds to an stoichiometric alloy with small concentration of
vacancies.  We obtain that, independently of $K$, the vacancies exhibit a
tendency to concentrate at the antiphase boundaries.  This effect gives rise,
via the vacancy-vacancy interaction (described by $K$), to an effective
interaction between bulk diffusing vacancies and moving interfaces which turns
out to strongly influence the domain growth process.  One distinguishes three
different behaviors:  $i)$ For $K/J<1$ the growing process of ordered domains is
anisotropic and can be described by algebraic laws with effective exponents
lower than $1/2$; $ii)$ $K/J \simeq 1$ corresponds to the standard Allen-Cahn
growth; $iii)$ for $K/J > 1$ we found that, although the motion of the interface
is curvature-driven, the repulsive effective interaction between both the
vacancies in the bulk and those at the interfaces provokes an slow down of the
growth.

\end{abstract}


\narrowtext

\section{Introduction}

The dynamical evolution of a binary alloy after a quench from a high temperature
disordered phase has been one of the prototypes in the study of relaxational
processes towards equilibrium.  It has been found that the late stages of this
process obey dynamical scaling and the typical domain size $R(t)$ dominates all
other lengths \cite{Komura88,Gunton83}.  In this regime, the domains grow in
time according to a power-law $R(t) \sim t^x$ with the growth exponent $x$
satisfying a remarkable universality.  For the case of a pure, ideal system, the
exponent $x=1/2$ \cite{Allen79} is associated with cases where the order
parameter is not conserved, whereas $x=1/3$ \cite{Lifshitz61} describes systems
with conserved order parameter.  For the non-conserved case a typical example is
a binary alloy undergoing an order-disorder phase transition.  Most theoretical
studies are limited to ideal conditions, that is to a pure, stoichiometric
binary alloy.  In such conditions, theory \cite{Allen79} and numerical
simulations \cite{Phani80,Sahni81,Fogedby88,Porta96} definitively agree about
the value of the kinetic exponent $x=1/2$.  Much less unanimity is obtained from
the experiments.  This is certainly due to the imperfections always present in
real materials.  Examples are:  vacancies, third-component impurities
\cite{Bley76}, non-stoichiometry \cite{Shannon88}, dislocations, etc...  It is
of great interest to elucidate in which manner and to what extent the presence
of these imperfections modifies the ideal asymptotic growth-law.

We shall not consider here the problem of quenched disorder \cite{Huse85}, but
concentrate on mobile punctual defects like vacancies, third-component
impurities and excess particles in off-stoichiometric binary alloys.  These
belong to the category commonly named annealed disorder.  Despite the
theoretical suggestion that such kind of disorder should not modify the
asymptotic growth-law \cite{Gilhoj95}, experiments \cite{Shannon88}, and
numerical simulation studies \cite{Gilhoj95,Srolovitz87,Shah90,Mouritsen90}
provide evidences for slow growth, either logarithmic growth laws or algebraic
laws with small exponents.

A general feature commonly observed during the early-time evolution in systems
with annealed defects is the tendency to concentrate the disorder at the domain
walls.  Indeed, experiments \cite{Shannon88,Oki77} and numerical simulations
\cite{Porta96,Srolovitz87,Shah90,Mouritsen90,Gilhoj95,Ohta87} reveal that the
vacancies and the excess particles tend to accumulate at the domain walls.  This
is accompanied by a depletion in the bulk defect concentration, that renders the
excess internal energy unsuitable to measure the total amount of interfaces.
Only at late times, as the interfaces disappear and the system approaches
equilibrium, the annealed defects may dissolve again into the bulk, provided
that they display no cooperative phenomena.  Simultaneously an overshooting in
the bulk order parameter is observed \cite{Gilhoj95,Gilhoj96}.  Very recently it
has been suggested \cite{Gilhoj95,Gilhoj95b,Gilhoj96} that this is a generic
effect in ordering dynamics, coming from a subtle competition between
non-equilibrium internal energy and non-equilibrium entropy.

In previous \cite{Srolovitz87,Shah90} studies of the diluted square Ising model
with nearest neighbors interactions, it was obtained that the effect of a small
concentration of vacancies is dramatic, leading to an extremely slow growth
described by a logarithmic growth-law \cite{Shah90} or even to a complete
pinning \cite{Srolovitz87}.  This difference, obtained on the same model in the
limit of low vacancy concentration, comes from special features introduced in
the coupled dynamics used; that is, the system evolves according to the
non-conserved Glauber dynamics but the vacancy concentration is forced to be
constant.  We notice that in these studies, the vacancies exhibit a natural
tendency to cluster.  The results obtained in the present work show that the
true asymptotic growth behavior is definitively algebraic with effective
exponents smaller than the Allen-Cahn value.

Concerning the present scenario for the effect of excess particles in a
non-stoichiometric binary alloys with no vacancies, very recently it has been
suggested \cite{Porta96} that the existence of effective interactions between
diffusing excess particles and those localized at the APB's is crucial in
determining the essential time dependence of the growth-law.  They showed that
when these specific interactions are not present the main assumptions underlying
the Allen-Cahn theory are fulfilled.  On the other hand there is experimental
evidence that small deviations from the stoichiometric composition may provoke
drastic modifications in the growth law.  It has been reported that the ordering
kinetics in $Cu_{0.79}Au_{0.21}$ \cite{Shannon88} shows a crossover from the
standard Allen-Cahn growth law, for stoichiometric $Cu_3Au$
\cite{Shannon88,Nagler88}, to a logarithmic growth-law.  In reference
\cite{Porta96} it was suggested that to account for such behavior, additional
interactions to the ones present in the (nearest neighbors) Ising model are
needed.  Indeed, the authors showed how this can be accomplished by simply
extending the interactions to next nearest neighbors.  Our main interest here
will be to incorporate the effect of such interactions between annealed
impurities in a more general framework.

The three-state Blume-Emery-Griffiths (BEG) model is specially adequate for our
purposes \cite{Blume71}.  For given values of the model parameters ($K$ and
$L$), the third value of the spin variable may represent either a vacancy or an
impurity (in particular an excess particle).  In the present work we take the
asymetric term $L=0$ and restrict to the case of vacancies in a stoichiometric
binary alloy, so that the additional coupling coming from the interplay between
the diffusive motion of both vacancies and excess particles is not considered
here.

The influence of mobile vacancies on the kinetics of ordering arises from the
coupling between their diffusive dynamics and the motion of the domain walls (in
this case, antiphase-domain boundaries (APB)).  This intercoupling depends on
the two facts concerning the behavior of the vacancies:  their tendency to
precipitate into the APBs and their tendency to cluster.  Whereas the former is
encountered for all values of the model parameters considered here , the
interaction among vacancies, and furthermore their tendency to cluster, depends
on $K$.  The combination of both effects gives rise to an effective interaction
(controlled by $K$) between bulk diffusing vacancies and those localized at the
interface that turns out to be crucial in determining the essential time
dependence of the growth-law.

The organization of the paper is the following.  We start by defining the model
and the region of parameters of interest here (section II).  In section III we
provide the details of the simulations and describe the algorithms used.  In
section IV we present the results and discuss them in section V.  Finally, in
section VI we summarize our main conclusions.

\section{The Model and Parameters}

We assume an underlying rigid square lattice with $i=1, \dots, N=l \times l$
sites that can be occupied either by $A$-atoms ($S_i=1$), $B$ atoms ($S_i=-1$)
or vacancies ($S_i= 0$).  Following standard procedures, the interactions are
taken to be pair-wise and restricted to nearest-neighbors (n.n.)  only.  The
spin-1 BEG hamiltonian is:

\begin{equation} {\cal H} = J \sum_{i,j}^{n.n.}  S_i S_j + K \sum_{i,j}^{n.n.}
S_i^2 S_j^2 + L \sum_{i,j}^{n.n.}  \left ( S_i^2 S_j + S_i S_j^2 \right )
\label{hamiltonian} \end{equation}
where, $J$ stands for the atom-atom ordering interaction, $K$ is a biquadratic 
coupling parameters accounting for the
energy difference between atom-atom pairs and those involving vacancies, and $L$
is an asymmetry term accounting for the energy difference between $A-A$ and
$B-B$ pairs.  When the parameter $K$ promotes the formation
of atom-atom pairs, the vacancies tend to cluster whereas if the vacancy-atom
pairs are prefered, cooperative effects for the vacancies are not expected, at
least in the limit of low vacancy concentration.

Although the hypothesis of a rigid lattice is crude, since lattice deformations
have an strong influence on the dynamics of the atoms around the vacancies, this
effect is partially taken into account in the phenomenological character
of the constant $K$.  We also expect that the assumption of pair interactions
only, disregarding many body effects, only introduces quantitative but not
qualitative changes concerning the ordering dynamics.

We have restricted the present study to concentrations so that the system
corresponds to a stoichiometric binary alloy with small concentration of
vacancies.  Since we have taken $J>0$, the ordered phase will be
antiferromagnetic-like, with almost all the bonds of the $A-B$ kind.

As we have mentioned in the introduction our goal here is to study the influence
of annealed vacancies on the kinetics of domain growth.  This influence
originates in the interplay between the two following specific interactions:  i)
the vacancy-APB interaction (here the APB is thought as a sequence of $A-A$ and
$B-B$ bonds) and ii) the vacancy-vacancy interaction.  When the the vacancy-APB
interaction is attractive, the vacancies tend to concentrate at the APBs and
therefore the vacancy-vacancy interaction introduces an effective interaction
between bulk vacancies and APBs.

Next step is to calculate these specific interactions for the different values
of the model parameters. These, obtained from the bond energies expressed
with respect the energy of the $A-B$ bond, are summarized in Table I.
For notation
we introduce the following reduced parameters $K^* = K/J$, $L^* = L/J$ , with
$J>0$.  It follows that $K^*=1$ separates the tendency for the vacancies to
cluster ($K^*<1$) or not ($K^*>1$).  Moreover, for $|L^*|<1$ the specific
interaction between vacancies and APB's is attractive, favoring the absorption
of vacancies at the interfaces.  For values of $|L^*|>1$ the behavior is more
complex, in particular, if $K^*<1$ a competition between $K^*$ and the
asymmetric term $L^*$ appears.

The results of the above analysis are illustrated in Fig.  \ref{FIG1}.  We have
indicated, in white, the region where the vacancies exhibit tendency to
accumulate in the APB's.  The line at $K^*=1$ separates the regions with vacancy
attraction and vacancy repulsion.  Black circles indicate the points studied in
the present work, all sitting along the line $L^*=0$.  The point at $K^*=0$ and
$L^*=0$ (indicated by an square) has been previously studied in\cite{Shah90} and
corresponds to a diluted Ising model.  The points with $K^*=1$ and $L^*= \pm 1$,
indicated by diamonds, correspond to a non-stoichiometric binary alloy without
vacancies \cite{Porta96}.

It is worth mention that the BEG-model, with $K=0$, $L=0$ and $J>0$, has also
been used for the study of the ordering dynamics via vacancies with $c_V<4.
10^{-4}$.  Such a restricted dynamics, only allowing exchanges between atoms and
vacancies may strongly modify the dynamics \cite{Vives92}.  Such mechanism is
not considered in the present study.

\section{Monte Carlo simulations}

We have performed different Monte Carlo simulations of the model defined in
section II with $L^*=0$, $J>0$ and $K^*$ inside de range $-0.5<K^*<1.4$.  First
we need to know, in the region of vacancy attraction ($K^*<1$), the temperature
at which the system separates into two phases.  This is important in order to
characterize the equilibrium state at the point to which quenches have been
performed.  Next, in order to characterize the time evolution of the ordering
process subsequent to the quench, we focus on the study of the structure factor
width.  It is known that the structure factor provides an overall description of
the ordering process.  In particular, the study of its time evolution will
provide information about the dynamical scaling properties.  These second group
of simulations constitute the major part of results presented and, contrarirly
to the equilibrium simulations, are very time consuming.

\subsection{\bf Equilibrium Simulations:  Phase Diagram.}

In order to obtain the phase diagram of model (1) in the particular case of
$L^*=0$, we have performed Monte Carlo simulations in the Grand Canonical
Ensemble using the Legendre transformation:

\begin{equation} {\cal H}_{GC} = {\cal H} - \mu \sum S_i^2 \end{equation}

Where $\mu$ stands for the chemical potential difference between atoms (either
$A$ or $B$) and vacancies.  This is because we restrict to the case of
stoichiometric composition ($N_A=N_B$) and the chemical potentials of $A$ and
$B$ are then equal.  The simulations have been performed on a system of linear
size $l=128$ using the Glauber dynamics implemented into the Metropolis
algorithm.  The different runs are extended up to $1500$ Monte Carlo steps per
site (MCs), our unit of time.

Figures \ref{FIG2}(a) and (b) show two different sections of the phase diagram.
Fig.  \ref{FIG2}(a) (Temperature $vs.$ vacancy concentration) corresponds to a
section with fixed $K^*=0$, and Fig.  \ref{FIG2}(b) (Temperature $vs.$ $K^*$) to
a section with fixed vacancy concentration $c_V=0.06$.  Phase $I$ corresponds to
an atomic disordered phase with randomly diluted vacancies; phase $II$
corresponds to an atomic ordered $AB$ phase with randomly diluted vacancies,
which exhibit only short range order; phase $III$ corresponds to a phase
separation region with coexisting ordered $AB$ domains with low concentration of
vacancies and vacancy clusters with a low concentration of disordered atoms.  In
order to limitate the range of model parameters to be explored in the Monte
Carlo simulations we first obtained a qualitative phase diagram using standard
Mean-field techniques.  For completeness, both results are simultaneously
presented in Fig.  \ref{FIG2}.

\subsection{Non Equilibrium simulations}

Although most of the results, presented in the next section, have been obtained
following the standard Kawasaki dynamics, alternative optimized algorithms have
been used when specially long simulations were needed.  This subsection is
devoted to the description of the different algorithms used in the study of the
time evolution of the process that follows a thermal quench from very high
temperature (disordered phase) to $T=0.1J/k_B$ performed on a stoichiometric
binary alloy with a small concentration of vacancies fixed at $c_V=0.06$ (being
$c_A=c_B=0.47$).  The different values of $K^*$ studied correspond to final
states into the ordered phases either $II$ and $III$.

\begin{enumerate}

\item{\bf Standard dynamical simulations.}  These are simulations performed
using the standard Metropolis algorithm together with the Kawasaki dynamics.
The linear system size is $l=200$ even though some initial studies where
performed on systems of $l=100$.  Starting from an initial disordered
configuration, the runs have been (typically) extended up to $20000 MCs$.
Moreover, averages over about 20 independent realizations have been performed.
From each simulation we have extracted the time evolution of the structure
factor defined as:

\begin{equation} S(\vec{k}) = \left | \frac{1}{N} \sum_{i=1,N} S_i \exp\left \{
{\rm i}\frac{2\pi}{a}\vec{k}\vec{r}_i \right \} \right |^2
\label{structurefactor} \end{equation}

where $\vec{k}$ are the reciprocal space vectors, $a$ is the lattice parameter
and $\vec{r}_i$ is the vector position of site $i$.  We have focused on the
profiles along the $(10)$ and $(11)$ directions around the superstructre peak at
$\vec{k} = (\frac{1}{2} \frac{1}{2})$.  Moreover, they have been averaged over
equivalent directions.  The size and shape of the ordered domains has been
obtained by fitting the averaged profiles to a lorentzian function powered to
$3/2$ in order to reproduce the Porod's law for the decay of the tail at long 
q's  \cite{Porods}:

\begin{equation} S(q,t)= \left \{ \frac{a(t)}{\displaystyle 1+ \left[
\frac{q}{\sigma_(t)}\right]^2} \right \}^{\frac32} \end{equation}

where $q$ is the distance to the superstructure peak
$q=|\vec{k}-(\frac{1}{2},\frac{1}{2})|$,
$a(t)$ is the height of the peak and $\sigma(t)$ the width.  Only data with
$S(q,t) > S_0 = 2.5 \; 10^{-5}$ has been considered for the fits.  $S_0$ has
been obtained from a completely disordered system with the same concentration of
particles and vacancies.  The quantities $a(t)$ and $\sigma(t)$ provide
information about the square order parameter growth and the inverse domains size
respectively.

\item{\bf N-fold way algorithm.}  For the particular case of $K^*=0$ and $L^*=0$
we have used the N-fold way algorithm \cite{Bortz75} in order to reach very long
times ($10^7 MCs$) in the evolution of a system of linear size $l=200$.  The
possible dynamic exchanges have been classified in 11 classes according to their
energy change.  (A class including the exchanges not affecting the system
configuration has also been taken into account, in order to compare with the
standard dynamical simulations in which time is spent in such exchanges).  The
time increment after each exchange has been taken as the expected time that, in
a standard Monte Carlo simulation, would pass until a useful proposed exchange
be accepted.  In this case, structure factor evolution has also been monitored,
as explained in the previous case. The averages have been performed over 
independent runs ($\sim 30$). 

\item{\bf Optimized multigrid algorithm.}  Given that for the more general case
of $K^*\neq 0$, the number of different energy classes is large, the N-fold way
algorithm is difficult to construct.  In this case, a less optimized algorithm
but simpler to implement has been constructed in order to reach very long times.
Starting from a standard multigrid algorithm \cite{Porta96,Shah90} we have made,
for each sublattice, a list of the exchanges whose probability of being accepted
is not negligible.  When a given sublattice is chosen, only the exchanges
present in the list are attempted.  The method turns out to be very efficient
when the number of attempted exchanges is low.  In particular, for the case
$K^*=0.6$ and $L^*=0$, we have performed simulations up to $10^7MCs$ with no
major difficulties (the linear system size was $l=200$ again). 

\end{enumerate}

\section{Domain growth results}

The Figure \ref{FIG3} shows snapshots of the microconfigurations as they evolve
with time after the quench for three different values of parameter $K^*$
(=$0.6$, $1.0$ and $1.4$).  The linear system size is $l=200$ and the quench
temperature $T=0.1 J/k_B$.  In this case we have used the standard dynamical
algorithm described before.  The vacancies are indicated in black while ordered
regions are incicated in white.  Due to the tendency exhibited by the vacancies
to concentrate at the interfaces the antiphase domain structure shows up
naturally.  The process of absorption of vacancies at the interfaces starts at
very early times and follows on until the whole interface network saturates.
This initial regime was studied previously \cite{Porta96} for the case of a
non-stoichiometric binary alloy.  Simultaneously, it was discussed in a more
general context and suggested to be a generic effect in ordering dynamics
\cite{Gilhoj95,Gilhoj95b,Gilhoj96}.  In any case, this is a transient, prior to
the long time domain growth regime of interest here.  Nevertheless, we remark
that this phenomenon of vacancy precipitation at the interfaces results of
crucial importance in the subsequent evolution (dictated by interface reduction)
specially when the interaction between vacancies is switched on ($K^* \neq 1$).
We now come back to Fig.  \ref{FIG3}.  Clear differences can be observed in
relation to both the orientation of the the antiphase boundaries and the speed
of the evolution towards equilibrium.  For $K^*=0.6$ (vacancy attraction), the
domains look square-like with the interfaces preferably directed along the (10)
direction.  In the case of $K^*=1.4$ (vacancy repulsion), although the domains
are square-shaped as well, the interfaces are directed along the (11) direction.
Moreover, in both cases, the interfaces tend to be flat (or almost flat), at
least in the regime depicted in Fig.  \ref{FIG3}.  As we shall discuss below,
this is a consequence of the vacancy-vacancy interaction which introduces energy
barriers for the motion of the vacancies localized at the interfaces favoring,
in each case, the different orientation of the APBs.  Contrarily, no preferred
orientation for the boundaries is observed when $K^*=1$.  Remember that in this
case there is no specific interaction between vacancies.  Concerning the speed
of the different evolutions, the fastest process occurs for $K^*=1$, whereas for
the other two cases it is clearly slower, apparently even more for $K^*=1.4$.
The introduction of specific interactions between vacancies seems to be behind
the slower evolutions, although the underlying physics is different.

Before proceeding further it is interesting to look at the quantitative results
obtained from structure factor calculations.  In Fig.  \ref{FIG4} we show the
time evolution peak width $\sigma (t)$ of the structure factor along the two
relevant directions $(10)$ (open circles) and $(11)$ (filled circles) for the
same three selected values of $K^*$ as in Fig.  \ref{FIG3}.  Dashed lines
indicate the regions of algebraic domain growth regime and the numbers on top
are the fitted values of the kinetic exponents.  We obtain that whereas for
$K^*=1$ and $K^*=1.4$ both, $\sigma_{(10)}$ and $\sigma_{(11)}$, evolve with the
same exponent, for $K^*=0.6$ the two exponents are clearly different.  This is
indicative of the existence of anisotropic growth and, as we shall discuss
later, it is related to the local accumulation of vacancies at the vertices of
the interfaces (see Fig.  \ref{FIG3}).  Moreover, while the values of the
exponents for $K^*=1.4$ and $K^*=0.6$ are definitively smaller than $1/2$, for
$K^*=1$ it is perfectly consistent with the standard Allen-Cahn value.  Notice
the large {\sl plateau} obtained in the case $K^*=1.4$.  One needs to perform
very large simulations before reaching the algebraic growth regime.  In fact, we
have also obtained such behavior for some values of $K^*<1$ inside the region
$-0.5<K^*<0$.  In addition, the extension of the {\sl plateau} depends on $K^*$
suggesting it is related to the existence of activated process with energy
barriers depending on $K^*$.  These energies increase (but not symmetrically) as
one varies $K^*$ from $K^*=1$, either to $K^*<1$ or to $K^*>1$, and in both
cases hinder the motion of the vacancies at the vertices of the interfaces.  The
expression for the associated energy barriers ($E_b^*=E_b/J$) are:  $E_b^* =
1-K^*$, for $K^*<1$ and $E_b^*= 3(K^*-1)$, for $K^*>1$.  In Fig.  \ref{FIG5}
black dots are the times needed to reach the algebraic regime for the different
values of $K^*$.  These have been estimated from the simulations as the ending
points of the {\sl plateau}.  Simultaneously we have plotted (discontinuous
lines) the passing-time, defined as $\tau \sim \exp(E_b^*/k_B T)$.  As an
example, we show the case of $K^*=0$ (Fig.  \ref{FIG6}).  The evolution up to
$\sim 10^7$ MCs has been obtained by following first the standard dynamical
simulations (circles) and next, by using the N-fold way algorithm (squares).
The arrow indicates the estimated time for the ending point of the {\sl
plateau}.

We have tested the existence of dynamical scaling.  Figures \ref{FIG7}(a), (b)
and (c) show the scaled structure factor profiles.  Those along direction $(10)$
have been shifted downwards four decades in order to clarify the picture.
Profiles along each direction have been conveniently scaled with the
corresponding $\sigma$.  The overlap of the data is satisfactory except for the
tails at large $q$'s.  Nevertheless, to prove the existence of dynamical scaling
it is necessary to have not only the collapse of the curves but also one
requires that both lengths, $\sigma_{(10)}$ and $\sigma_{(11)}$, evolve with the
same power-law.  This happens to be the case for $K^*=1$ and $K^*=1.4$ whereas
for $K^*=0.6$ both sets of profiles scale independently according to widths
evolving with different power-laws.

In the following two figures we present a complete study of both the anisotropic
character of the growth and the kinetic growth exponent(s) for a wide range of
values of the interaction parameter $K^*$.  Fig.  \ref{FIG8} shows the time
evolution of the ratio $\eta\equiv \sigma_{(11)}/\sigma_{(10)}$ for different
values of $K^*$.  For $K^*<0.8$, $\eta$ definitively increases with time,
showing that the shape of the ordered domains becomes more and more spike-like.
For $0.8<K^*<1.0$ the ratio $\eta$ remains constant indicating that the domains
are circular during all the evolution.  For $K^*>1$ after an initial decrease,
$\eta$ reaches the value $1/\sqrt 2$ indicating that, at large times, the
domains are square-like and grow isotropically.

Figure \ref{FIG9} shows the growth exponents obtained by fitting an algebraic
growth law to the evolution of both $\sigma_{(10)}$ (open circles) and
$\sigma_{(11)}$ (filled circles).  Note that Allen-Cahn values ($n\simeq 0.5$)
are only reached for values of $K^*\sim 1$.

We have also studied the behavior of the structure factor at large $q's$.  They
are affected by two different phenomena.  (i) Along the direction $(1,1)$ for $q
> 0.5$ the structure factor is distorted due to the existence of the non-scaling
fundamental peak at $\vec{k}=(00)$.  (Strictly speaking the value of the
structure factor at $\vec{k}=(00)$ is always zero, but the peak exhibits some
finite width due to the existence of disorder in the system) (ii) For the cases
in which the vacancies dissolve into the bulk, there is an homogeneous
non-scaling background which, in turn, may evolve in time.

\section{Discussion}

The differences in the values obtained for the kinetic exponent of the growth
law lie on the different characteristics of the intercoupling between bulk
diffusing vacancies and the interface motion.  For $K^* \neq 1$ this
intercoupling proceeds via an effective interaction originated from the
vacancy-vacancy specific interaction.  Moreover, this introduces differences in
the internal structure of the interface.  In the particular case of $K^*=1$,
this intercoupling reduces to a simple encounter between curvature-driven
interfaces and mobile vacancies which mutually cross their respective
trajectories.  This does not make the curvature ineffective but may slow down
the domain growth.  It has been shown that \cite{Porta96} the effect of this
simple intercoupling does not modifies the essential time dependence of the
growth law but modifies the growth rate (prefactor) which decreases as the
mobile impurity concentration increases.

We next discuss separately the other two cases.  We start with the case of
vacancy attraction ($K^*=0.6$) and point out some other relevant features
present in Fig.  \ref{FIG3}.  Notice the increasing concentration of vacancies
at the interfaces as they evolve with time.  This is premonitory of the phase
separation process, eventually reachable at longer times.  During the regime
showed in Fig.  \ref{FIG3} one is mainly concerned with a non homogeneous
distribution of vacancies along the interfaces.  Special mention deserves the
local accumulation at the vertices.  This is the signature of a previous fast
process of interface reduction (due to the high-curvature of the vertices).  The
further evolution is hindered until the vacancies diffuse along the interfaces.
This involves activated processes.  Indeed, in our simulations we have observed
how the temporal pinning of the high-curvature portion of the interfaces
provokes that the further evolution of the interconnecting interfaces (with
lower curvature) does not fulfill the main assumptions underlying the Allen-Cahn
theory.  The importance of this temporal pinning depends on $K^*$, since the
energy barriers ($E_b^*=1-K^*$) for a vacancy at the corner of the interface to
move increases as $K^*$ decreases from $K^*=1$.  In Fig.  \ref{FIG10} we show
this effect for the case $K^*=0$.  In particular, one observes how the single
(rectangular) domains evolve so that they become more and more plate-like.  This
is because the longer interfaces (with a lower concentration of vacancies) are
the only able to evolve.  Moreover, they remain (almost) flat and parallel to
the (10) direction.  This feature is not observed in previous studies on the
diluted antiferromagnetic Ising model \cite{Shah90} (notice that it corresponds
to the BEG with $K^*=0$) so that it cannot be attributed exclusively to the
interaction (which favors the vacancies to be nearest neighbors at the
interface).  In Fig.  \ref{FIG11} we schematically illustrate this mechanism of
evolution.  It represents a typical domain in the regime under discussion here;
that is, the regime characterized because the width of the interface is small
and the only relevant length is the size of the AB ordered domains.  Two facts
have to be taken into account:  the attractive vacancy-vacancy interaction
defined by the hamiltonian and the conserved character of the implemented
dynamics (Kawasaki).  The combination of both introduce energy barriers for the
motion of the vacancies (circles) at the interfaces which hinder the shrinkage
of the domain.  We have indicated in black (white) the vacancies with associated
energy barriers so that they are induced to go outwards (inwards).  The
vacancies in grey do not have any preference.  The crucial point is that the
motion inwards of the vacancies at the corner are strongly hindered whereas for
its neighbors, it is favored.  This provokes that the shrinkage of the domains
proceeds by displacing the flat interfaces and accumulating the excess vacancies
at the vertices\cite{flat}.  The domains become then spike-like along the (11)
direction breaking down the single-length dynamical scaling and so the growth is
anisotropic.  In reference \cite{Shah90}, the authors coupled the Glauber
dynamics for the spins to the conserved dynamics for the number of vacancies in
such a way that the vacancy at the corner is not pinned.  Nevertheless they
found that the dynamical evolution of the ordering process is effectively
described by a logarithmic growth-law (in the limit of low vacancy
concentration).  Other studies \cite{Srolovitz87} on the diluted ferromagnetic
Ising model, encountered that, by coupling both dynamics differently (the
simultaneous vacancy-spin exchange and spin flip is not allowed) so that the
vacancy at the
corner is effectively pinned, the growth stops.  In the case of the alloy, the
dynamics implemented follows directly from the requirement of the conservation
law for the number of particles and we found that the ordering process is
definitively described by a growth-law that although being slower than the
Allen-Cahn law it is definitively algebraic.  For some values of $K^*$ this
algebraic regime is preceded by a {\sl plateau} with extension depending on
$K^*$.  More precisely, it is larger as bigger the energy barriers are.  In
particular, we found that for $K^*=0$ this algebraic regime is visible only
after $\sim 10^5$ MCs (see Fig.  \ref{FIG6}).  We notice that the BEG model, in
the particular case of $K=L=0$, corresponds to a diluted Ising model.  In view
of this, we believe that the growth-law for the diluted antiferromagnet is
algebraic.  In fact, the authors in \cite{Shah90} did not excluded this
possibility in their discussion.  Concerning, the complete pining of the process
reported by \cite{Srolovitz87}, it is certainly due to the extremely low
temperature.

Later on, as the width of the interface increases, this pinning effect,
localized at the corner, becomes less important and one expects the system
crosses over to a complete different regime.  In this asymptotic regime, not
reached in our simulations, to talk about the interface, as formed by the
increasing accumulation of vacancies, is meaningless.  Rather, one would deal
with a phase-separation process which is not studied in the present work.

We now focus on the discussion for $K^*=1.4$.  In this case, the repulsive
interaction between vacancies favors them to be next nearest neighbors at the
interfaces.  As previously, the driving force is contained in the corner but its
motion is hindered by a barrier of energy $2(K^*-1)$.  The interface connecting
two vertices is directed along the (11) direction and evolves so that it
displaces parallelly.  The energy barrier associated to this mechanism is
$E_b^*=3(K^*-1)$.  The vacancies, in excess as a consequence of the interface
reduction, are in this case (vacancy repulsion) expelled to the bulk.  Moreover,
the migrating interface has to cope with the effective repulsive interaction
with the bulk vacancies, which concentration increases as the system evolves.
Indeed, one expects this should interfere the dynamics.  At this point, it is
interesting to discriminate whether or not this interference makes the
curvature-driven mechanism to become ineffective.  In this sense, we have
verified that the interface evolves covering a domain area constant in time.
This is shown if Fig.  \ref{FIG12} for single domains directly extracted from
our simulations.  These calculations have been performed using the optimized
multogrid algorithm discussed previously.  A linear time dependence for the
domain area evolution is obtained.  This is commonly accepted to be indicative
that the motion of the interface is curvature-driven \cite{Domany90}.  It then
follows that the effect of the intercoupling between mobile bulk vacancies and
the evolving interfaces is, in this case, to slow down the global process as it
is revealed by a decreasing in the effective growth exponent but it does not
make ineffective the curvature driven mechanism which remains as the underlying
mechanism for the motion of the interfaces.  Moreover the effective exponent
tends to the ideal $1/2$-Allen-Cahn value as $K^* \rightarrow 1$.

\section{Summary}

We use a Blume-Emery-Griffiths model (with $L=0$) to study the influence of
mobile vacancies in the kinetics of domain growth in a stoichiometric binary
alloy after quenches to very a low temperature ($T=0.1J/k_B$) through an
order-disorder transition.  The study is performed by Monte Carlo simulations in
the limit of low vacancy concentration for a wide range of values of the
biquadratic coupling parameter $K^*$ which controlles the specific
vacancy-vacancy interaction.  For all values of $K^*$ inside the range
$-0.5<K^*<1.4$ we found that the vacancies tend to concentrate at the
interfaces.  This feature introduces, via the parameter $K^*$, an intercoupling
between diffusing bulk vacancies and moving interfaces.  In the particular case
of $K^*=1$ this intercoupling does not take place via any specific interaction
and the ordering process is consistent with the Allen-Cahn law.  In fact we find
that $n \sim 1/2$ for $K^* \sim 1$ whereas the exponent is clearly smaller when
such interaction is present, no matter if it is attractive ($K^*<1$) or
repulsive ($K^*>1$).  Nevertheless, our results clearly show that the growth is
definitively algebraic.  When $K^*<1$ the attractive vacancy interaction favours
the increasing accumulation of vacancies at the interfaces as the system
evolves.  The regime of interest here corresponds to the very initial stages of
a phase separation process when the width of the interfaces is small and the
only relevant length is the size of the $AB$ ordered domains.  Our main finding
is that for quenches inside the coexistence region the growth for the binary
alloy is, in this regime, anisotropic.  This is related to the existence of
energy barriers (depending on $K^*$) which hinder the motion of the vacancies at
the vertices of the (10) square-like domains and simultaneously provoke local
accumulations of vacancies along the (11) directions.  Furthermore, these
barriers may delay the apparition of the algebraic regime of the growth process
to very late times.  Concerning the underlying mechanism for the motion of the
interfaces, it is not purely curvature driven and the two effective exponents
nedded to describe the process are lower than the Allen-Cahn value.
Nevertheless, both tend to $n=1/2$ as $K^* \rightarrow 1$.  For $K^*>1$ the
specific vacancy-vacancy interaction is repulsive and the interfaces of the
square-like domains are in this case directed along the (11) directions.  As in
the previous case, the motion of the vacancies at the vertices is an activated
process.  The associated energy barriers depend on $K^*$ and the algebaric
growth regime shows up only after the time needed for overpassing the barrier.
Since the interface motion has to cope with the repulsive diffusive vacancy
interaction, we found that the ordering process clearly slows down.
Nevertheless, this repulsion does not make the curvature ineffective.  The
effective exponent is lower than the Allen-Cahn value but approaches to $n=1/2$
as $K^* \rightarrow 1$, as it is expected.

\acknowledgements We acknowledge financial support from the Comisi\'on
Interministerial de Ciencia y Tecnolog\'{\i}a (CICyT, project number MAT95-504)
and supercomputing support from Fundaci\'o Catalana per a la Recerca (F.C.R.)
and Centre de Supercomputaci\'o de Catalunya (CESCA).  M.P.  and C.F.  also
acknowledge financial support from the Comissionat per a Universitats i Recerca
(Generalitat de Catalunya).

\newpage

\begin{table} \caption{Bond energies and the same measured with respect to the
$A-B$ bond for the BEG model as a function of the parameters $J$, $K$, and $L$.}
\begin{tabular}{ccc} Bond & Energy & Excess energy \\ \hline $A-B$ & $-J+K$ & 0
\\ $A-A$ & $J+K+2L$ & $2J+2L$ \\ $B-B$ & $J+K-2L$ & $2J-2L$ \\ $A-V$ & $0$ &
$J-K$ \\ $B-V$ & $0$ & $J-K$ \\ $V-V$ & $0$ & $J-K$ \end{tabular} \label{TAB1}
\end{table}

\newpage

%%%% Fig01

\begin{figure} \caption{Regions in the space of the parameters $K^*$ and $L^*$
where different dynamics are expected.  The solid line $K^*=1$ separates the
region of vacancy attraction $(K^*<1)$ from the region of vacancy repulsion
$(K^*>1)$.  The solid lines $L^*=\pm 1$ separates the region where the vacancies
tend to precipitate at the antiphase boundaries $(|L^*|<1)$ from the region of
vacancy-antiphase boundary repulsion $(|L^*|>1)$.  Solid circles indicate the
points studied in the present work.  The square corresponds to the case
$K^*=L^*=0$ and the diamonds are the points where the model is formally
equivalent to a non-stoichiometric binary alloy.}  \label{FIG1} \end{figure}

%%%% Fig02

\begin{figure} \caption{(a) $T$ $vs.$ $c_V$ phase diagram of the BEG model for 
$K^*=L^*=0$.  (b) $T$ $vs.$ $K^*$ phase diagram of the BEG model for $L^*=0$ and
$c_V=0.06$. The points correspond to the Monte Carlo results whereas the solid 
lines are obtained from mean-field calculations. Dashed lines are just guides to 
the eyes. In the insert, we show the region of interest. The arrows indicate the 
working temperature and vacancy concentration}  \label{FIG2} \end{figure}

%%%% Fig03

\begin{figure} \caption{Snapshots of the evolving domain structure for
$K^*=0.6$, $K^*=1.0$ and $K^*=1.4$.  Vacancies ($A$ and $B$ particles) are
painted in black (white).  The simulations are performed in a 200$\times$200
square lattice with $c_V=0.06$ at $T=0.1J/k_B$.}  \label{FIG3} \end{figure}

%%%% Fig04

\begin{figure} \caption{Width of the structure factor, $\sigma$, $vs.$ time for
$K^*=0.6$, $K^*=1.0$ and $K^*=1.4$.  Open circles correspond to the $(10)$
direction and filled circles to the $(11)$ direction.  Dashed lines indicate
the regions where the growth exponents, written on top, are
fitted.}  \label{FIG4} \end{figure}

%%%% Fig05

\begin{figure} \caption{Time needed to reach the algebraic regime $vs.$ $K^*$
(black circles).  Dashed lines are the predicted slopes of these curves from the
energy barriers present in the model.}  \label{FIG5} \end{figure}

%%%% Fig06

\begin{figure} \caption{Structure factor width, $\sigma$, $vs.$ time for
$K^*=0$.  Open/filled symbols correspond to the (10)/(11) direction.  The end
of the {\sl plateau} is taken to be at the inflection point, as indicated by the
arrow.  The results shown by circles have been obtained following standard 
simulations whereas the long time results, indicated by squares, have been 
obtained by using the N-fold way algorithm. The system size is 200$\times$200,  
$c_V=0.06$ and $T=0.1J/k_B$.}  \label{FIG6} \end{figure}

%%%% Fig07

\begin{figure} \caption{log-log plot of the scaled structure factor profiles in
the $(10)$ and $(11)$ directions for $K^*=0.6$, $K^*=1.0$ and $K^*=1.4$.  The
profiles in the (10) direction have been shifted four decades below in order to
clarify the picture.}  \label{FIG7} \end{figure}

%%%% Fig08

\begin{figure} \caption{Ratio $\eta \equiv \sigma_{(10)}/\sigma_{(11)}$ $vs.$
time for different values of the parameter $K^*$.  All these simulations are
performed in a 200$\times$200 square lattice with $c_V=0.06$ at $T=0.1J/k_B$.}
\label{FIG8} \end{figure}

%%%% Fig09

\begin{figure} \caption{Growth exponent, $vs.$ $K^*$ obtained from the structure
factor width, $\sigma$.  Open (filled) circles correspond to the evolution of
$\sigma_{(10)} (\sigma_{(11)}).$ All the simulations are performed in a
200$\times$200 square lattice with $c_V=0.06$ at $T=0.1J/k_B$.}  \label{FIG9}
\end{figure}

%%%% Fig10

\begin{figure} \caption{Snapshots of some evolving domains directly extracted
from our simulations for $K^*=0$.  Vacancies ($A$ and $B$ particles) are painted
in black (white).  The simulations are performed in a 200$\times$200 square
lattice with $c_V=0.06$ at $T=0.1J/k_B$.}  \label{FIG10} \end{figure}

%%%% Fig11

\begin{figure} \caption{Schematic representation of the evolution of a square
domain typically observed in the $K^*<1$ simulations.  The circles represent
vacancies whereas the white (black) squares represent $A (B)$ particles.  The
mechanism of evolution is emphasized by painting in black (white) those
vacancies which have the tendency to move outwards (inwards).  The vacancies
painted in grey are the ones which are, in this sense, indifferent.}
\label{FIG11} \end{figure}

%%%% Fig12

\begin{figure} \caption{Domain area $vs.$ time of a single domain for $K^*$=1.2
(solid and dotted lines) and $K^*=1.4$ (dashed line).  The solid (dotted) line
corresponds to $c_V$=0.06 ($c_V$=0.04).  We simultaneously show, in the inset,
the time evolution of the total amount of bulk vacancies in the system.  The
simulations are performed in a 200$\times$200 square lattice at $T=0.1J/k_B$.}
\label{FIG12} \end{figure}


\begin{references}
\bibitem{flat} If this analysis is done in a circular domain, it can be seen
that it first evolves to a square domain and then proceeds as discussed.
\end{references}
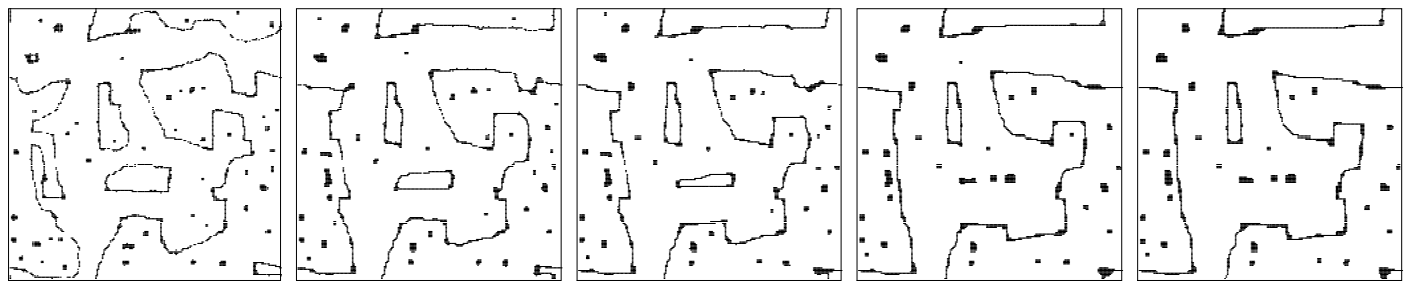
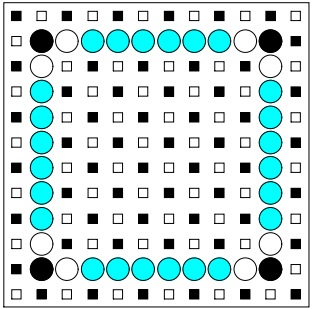
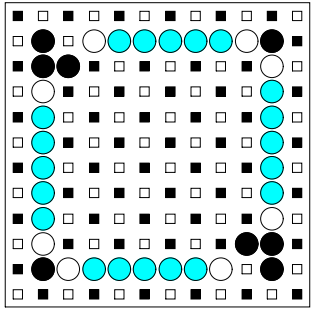
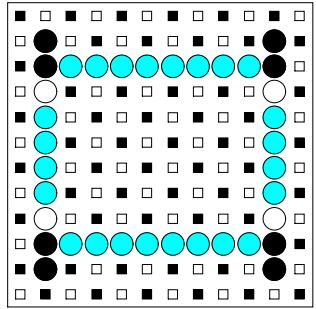
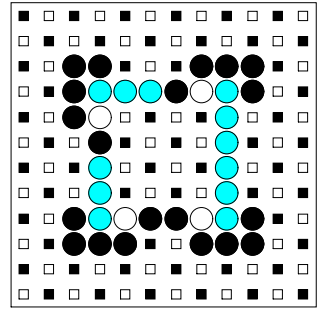
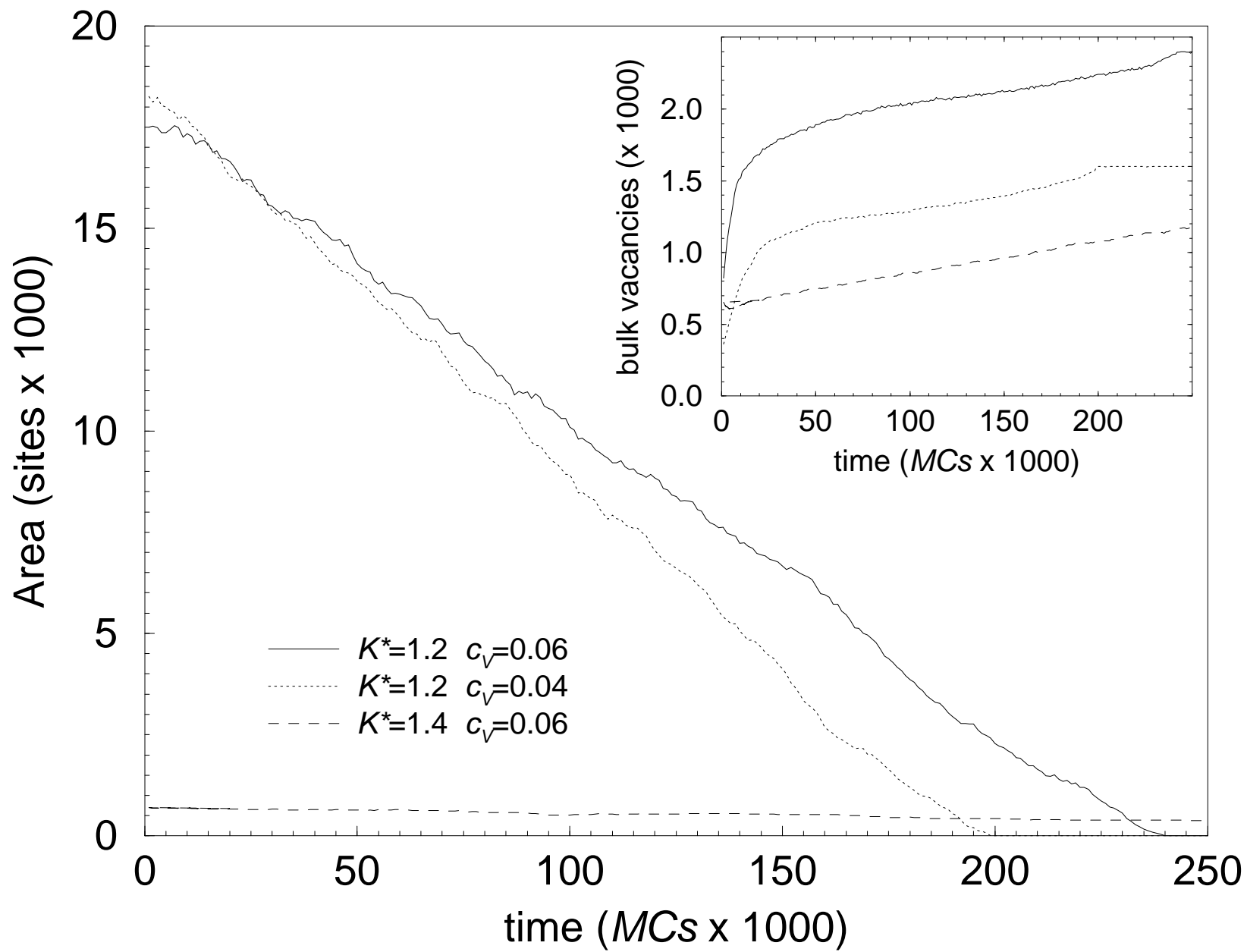
\end{document}